\documentclass[11pt]{article}
\usepackage{hyperref}
\pdfoutput=1

\begin{document}

\title{Vapor bubbles 'SUMO'}

\author{Ryota Hosoya$^{1}$ \& Ichiro Ueno$^{2}$ \\
\\\vspace{0pt} $^{1}$Graduate School, Tokyo University of Science, \\\vspace{6pt} 2641 Yamazaki, Noda, Chiba 278-8510, JAPAN
\\\vspace{6pt} $^{2}$Tokyo University of Science, Noda, JAPAN}

\maketitle


\begin{abstract}
We introduce growing and condensing processes of vapor bubble(s) injected into a subcooled pool in this fluid dynamics video. 

\end{abstract}

\section{Introduction}



We carry out a series of experiments with a special interest on growing and condensing processes of vapor bubble(s) injected into a subcooled pool. We examine effects of the degree of subcooling of the bulk in the pool and injection rate of the vapor into the pool. We pay our special attention to (i) abrupt collapse of the injected vapor bubble to form micrometer-scale bubbles, and (ii) interaction of adjacent vapor bubbles laterally injected to the pool through the orifices. We have found that a fine disturbance does arise on the surface of the vapor bubble just prior to its abrupt collapse. The bubble never exhibits an abrupt collapse without such instability over the free surface. In the case of the injection of a pair of vapor bubbles through the neighboring orifices, the interaction between the bubbles and the effects of the induced flows by the bubble behaviors control the surface dynamics. This fluid dynamics video introduces those phenomena.

%
\end{document}